# $Bi_2Te_3$–$Sb_2Te_3$–$Bi_2Te_3$ Lateral Heterostructures Grown by Molecular Beam Epitaxy

Puspendu Guha,[1,2,*] Sangmin Lee,[3] Eunsu Lee,[1] Hyeonhu Bae,[4] Hoonkyung Lee,[4] Miyoung Kim[3] and Gyu-Chul Yi[1,*]

[1]Department of Physics and Astronomy, Institute of Applied Physics, Seoul National University, Seoul 08826, Republic of Korea

[2]International Iberian Nanotechnology Laboratory (INL), Avenida Mestre José Veiga s/n, Braga 4715-330, Portugal

[3]Department of Materials Science and Engineering and Research Institute of Advanced Materials, Seoul National University, Seoul 08826, Republic of Korea

[4]Department of Physics, Konkuk University, Seoul 05029, Republic of Korea

***Email:** puspendu.guha@inl.int and gcyi@snu.ac.kr

## Abstract

Lateral in-plane heterostructures enable precise control of electronic properties and quantum effects in 2D materials. However, their periodic synthesis is challenging because it requires precise control to maintain sharp, coherent interfaces and compatible growth conditions across different domains. Herein, we report the successful heteroepitaxial growth of $Bi_2Te_3$–$Sb_2Te_3$–$Bi_2Te_3$ and periodic lateral heterostructures on hexagonal boron nitride (hBN) through *in-situ* multiple growth steps at different stages using a molecular beam epitaxy (MBE) system. These trilateral heterostructures are fabricated by growing triangular or hexagonal $Bi_2Te_3$ islands at the very beginning, with typical sizes of several hundred nanometers, on the single-crystalline hBN, followed by the lateral growth of $Sb_2Te_3$ to form bilateral heterostructures, and finally growing $Bi_2Te_3$ on the side facets of the bilateral heterostructures. The electron microscopy results confirm the core area as $Bi_2Te_3$, the intermediate layer as $Sb_2Te_3$, and the outermost region as $Bi_2Te_3$. The resulting heterostructures are approximately 4–8 nm thick and several hundred nanometers in lateral dimensions. These heterostructures are found to grow epitaxially on hBN (< ±4° misalignment), and the individual layers are strongly epitaxially aligned with each other. The in-plane heterojunctions are analyzed using the aberration-corrected ($C_s$-corrected) high-angle annular dark-field scanning transmission electron microscopy technique. We have explored and established the plasmonic properties of these fabricated $Bi_2Te_3$–$Sb_2Te_3$–$Bi_2Te_3$ lateral heterostructures. In addition, the electronic states and the topological properties of the few quintuple layers (QLs) (2- to 4-QLs) $Bi_2Te_3$–$Sb_2Te_3$ lateral periodic heterostructures are investigated by first-principles calculations.

## 1. Introduction

Topological insulators (TIs) are the new states of quantum matter. Their band structures comprise a bulk energy gap but linearly dispersed surface states of the exotic gapless edge with spin-orbit interactions and time-reversal symmetry.[1-3] Over the past decade, group V-VI TI materials, such as $Bi_2Se_3$, $Bi_2Te_3$, and $Sb_2Te_3$, have garnered significant interest because of their exotic physical and chemical properties.[1-4] They all share identical rhombohedral crystal structures [space group $D_{3d}^5 (R\overline{3}m)$], with repeating units of five atomic Te–Bi–Te–Bi–Te layers [for example $Bi_2Te_3$], called quintuple layers (QLs) (with approximately 1 nm thickness) stacked via van der Waals (vdW) interactions.[1,4,5]

Lateral heterostructures formed by in-plane bonding of van der Waals (vdWs) materials have emerged as a key direction for band engineering, interface physics, and device miniaturization.[6-11] The lateral junctions provide a pathway to integrate the spin and charge transport properties of TIs, a very intriguing subject matter in condensed matter physics.[12-14] Owing to strong in-plane covalent bonding at interfaces, advanced/complex TI-based configurations can exhibit tunable electronic, optical, spintronic, and structural properties, offering new routes for device applications.[12-16] In particular, multilateral heterostructures, such as p–n–p or n–p–n junctions, are very promising. Such architectures mimic classical bipolar junction transistors (BJTs), essential components in modern electronics. [16,17] Extending this concept to 2D TIs can enable novel topological BJT architectures, where charge and spin transport are both tunable. However, accomplishing the multilateral junctions is a very challenging and complex matter to date.[17,18] It is even more difficult and troublesome to fabricate them in a controlled way with atomically clean and clear interfaces and single-crystalline individual components,[18,19] which are necessary for devices with low noise.[20] On the other side, strong covalent bonds in lateral heterostructures create in-plane interfacial states that enable innovative configurations and can lead to new optical, electronic, and quantum phenomena through interface-driven modifications.[6,18,21-23]

To date, several reports have been published on various (bi) lateral heterostructures.[19,24-28] However, controlled fabrication of these heterostructures remains a substantial challenge.[18,25,16] Thus, reliable and successful fabrication of trilateral or multilateral TIs remains scarce and difficult.[18,29] A few reports on advanced or complex lateral heterostructures have been reported so far; however, they are limited to transition metal dichalcogenides (TMDs)-based heterojunctions.[18,23,29-34]

Very recently, Goyal et al. demonstrated a one-pot, defect-mediated synthesis of lateral $Bi_2Te_3$–$Sb_2Te_3$–$Bi_2Te_3$ heterostructures, but the method lacks defect-free and purely lateral growth. Our group reported the heteroepitaxial lateral growth of $Bi_2Te_3$–$Sb_2Te_3$ using the molecular beam epitaxy (MBE) technique.[5] These works open the possibility of expanding binary TI junctions into complex multilateral structures. However, one major challenge lies in preventing unwanted alloy formation between constituents during the lateral stitching process[35,36] and, thus, has not been realized so far. Therefore, fabricating advanced lateral heterostructures requires a precise and controllable synthesis technique with finely tuned growth parameters to maintain phase purity and sharp in-plane interfaces.

To meet these requirements, MBE offers a significant advantage. It enables spatially uniform, atomically controlled, high-quality growth suitable for high-purity heterostructures with precise spatial resolution.[4-5] This work communicates a novel MBE-based synthesis of $Bi_2Te_3$–$Sb_2Te_3$–$Bi_2Te_3$ trilateral and periodic lateral heterostructures on single-crystalline hBN substrates. $Bi_2Te_3$ and $Sb_2Te_3$ are selected as *n-type* and *p-type* materials, respectively,[36] forming an n–p–n configuration. To our knowledge, this paper reports an initial demonstration of the successful *in-situ* synthesis of thin-layered trilateral and multilateral TIs heteroepitaxial structures. Our results highlight sharp interfaces, epitaxial alignment, and promising electronic and plasmonic properties of lateral TI heterostructures.

## 2. Results and Discussion

We executed a multi-step *in-situ* growth process to engineer trilateral heterostructures on a 2D material (hBN) under UHV conditions [chamber base pressure: ~$10^{-9}$ torr] (as illustrated in Figure 1). First, faceted $Bi_2Te_3$ islands with typical lateral sizes of several hundred nanometers were grown on a single-crystalline hBN template via a two-step growth process (Figure 2(a)). Subsequently, bilateral $Sb_2Te_3$–$Bi_2Te_3$ heterostructures were formed in a single-step by slightly lowering the substrate temperature[5] [see Figure 2(b)]. In the final stage, $Bi_2Te_3$ was reintroduced through another single-step process at further reduced temperature to fabricate Bi$_2$Te$_3$–Sb$_2$Te$_3$–Bi$_2$Te$_3$ trilateral heterostructures. To suppress alloying between Bi and Sb, each layer was grown sequentially at different temperatures, with each successive layer deposited at a lower temperature. In order to achieve multilateral heterogeneous epitaxial structures successfully, the second material needs to be nucleated successively at the edge of the first material, followed by the third material nucleating at the edge of the second layer, and so on [depicted in Figure 1(c)]. The edge atoms of each preceding material act as the nucleation sites for the subsequent layer.[5,9,18,23,37] In this work, an *in-situ* growth strategy was crucial to obtain trilateral and multilateral heterostructures; the edge atoms would otherwise become terminated or passivated, if not devised in an *in-situ* process.[37]

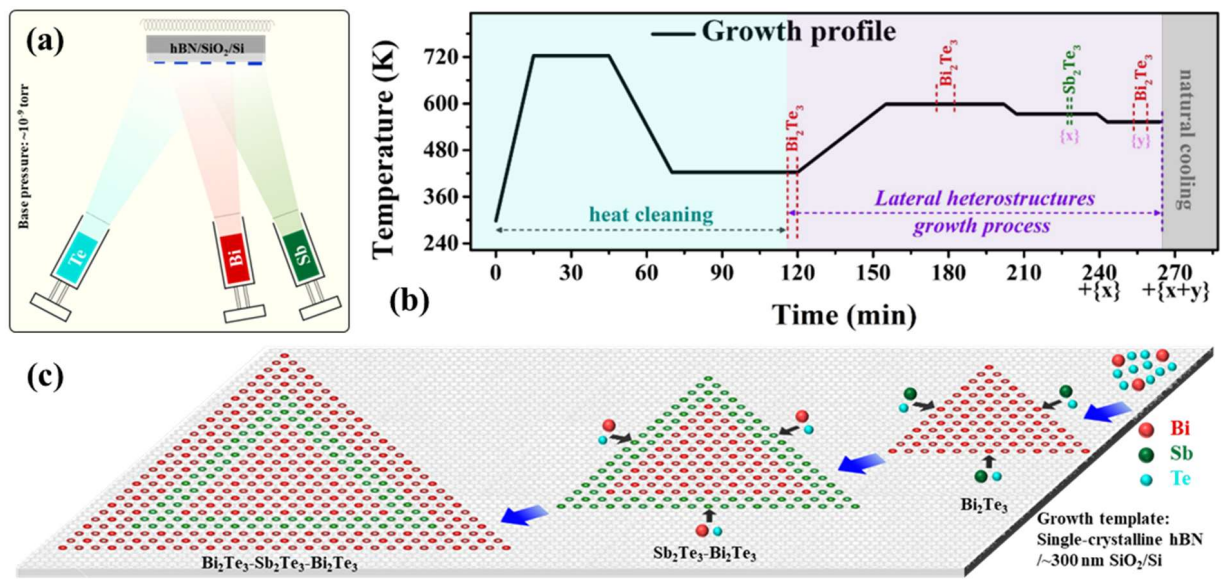

**Figure 1.** (a) Schematic of the growth chamber setup. (b) Growth profile and (c) illustration of Bi$_2$Te$_3$–Sb$_2$Te$_3$–Bi$_2$Te$_3$ lateral heterostructure on single-crystalline hBN/~300 nm SiO$_2$/Si.

The low- and high-magnification FESEM micrographs of the as-fabricated structures are depicted in Figures 2(c-d); the inset of Figure 2(d) presents a single as-prepared trilateral heterostructure. Large-scale spatially uniform trilateral heterostructures on hBN can be recognized in Figure 2(c). Figure 2(d) and its inset show different lateral contrasts in the SEM images, implying the successful synthesis of the trilateral heterostructures. The brighter contrasts at the core and the third layer correspond to Bi$_2$Te$_3$, whereas the darker contrast (intermediate layer) stands for Sb$_2$Te$_3$. The pure lateral growth is discussed in the following section. The surface morphologies of the as-grown trilateral heterostructures were further investigated by AFM to obtain surface and thickness information of the as-grown trilateral heterostructures. Figure 2(e) shows a representative AFM image of the as-fabricated heterostructures on hBN and the thickness profile (~ 6 nm) of a single heterostructure obtained across the marked line. The overall thicknesses of these trilateral heterostructures were found to be in the range of 4–8 nm.

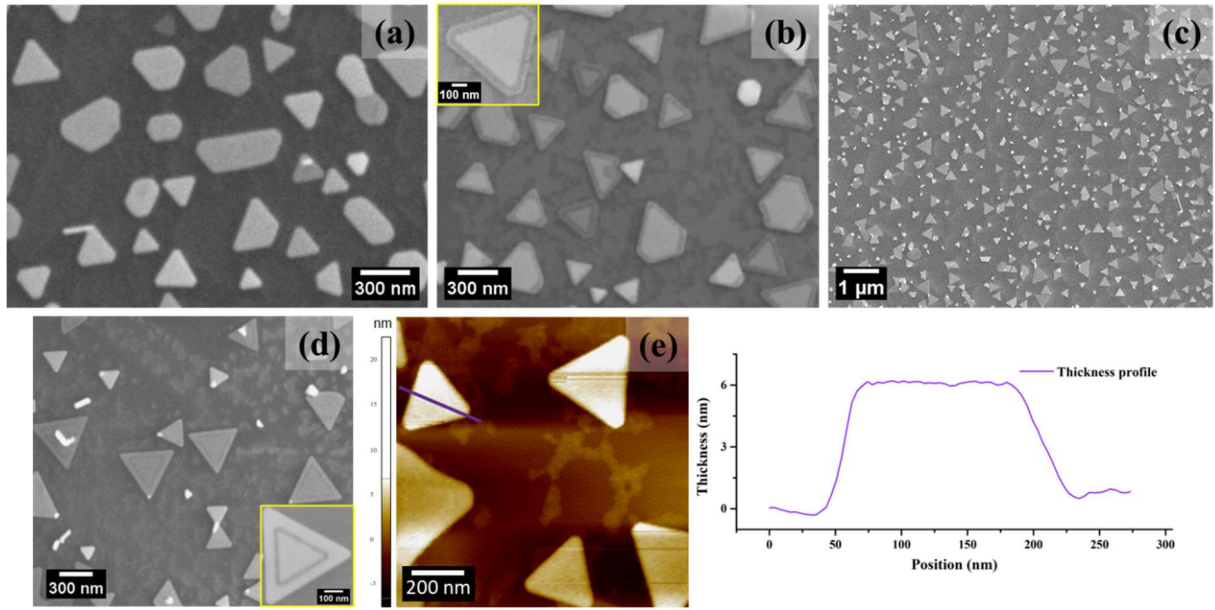

**Figure 2.** FESEM images of the as-grown faceted (a) $Bi_2Te_3$ islands, (b) $Sb_2Te_3$–$Bi_2Te_3$, and (c-d) $Bi_2Te_3$–$Sb_2Te_3$–$Bi_2Te_3$ lateral heterostructures; insets of (b) and (d) show the corresponding single heterostructures. (e) AFM image of the trilateral heterostructure and the corresponding thickness profile obtained across the marked line.

The structural properties of the $Bi_2Te_3$–$Sb_2Te_3$–$Bi_2Te_3$ lateral heterostructures were investigated using TEM and allied techniques. Figure 3(a) shows a low-magnification bright-field (BF)-TEM plan-view image of the trilateral heterostructures on hBN. The selected area electron diffraction (SAED) pattern, obtained from a single heterostructure (shown in the inset of (a)) along the [0001] zone axis, is presented in Figure 3(b). The diffraction pattern reveals the hexagonal symmetry of both the single-crystalline hBN and the as-fabricated structure. The measured angle between $\{10\text{-}10\}_{hBN}$ and $\{11\text{-}20\}_{structure}$ was approximately 30°, implying $\{10\text{-}10\}_{hBN} \parallel \{10\text{-}10\}_{structure}$. This indicates an epitaxial footprint between hBN and the as-grown lateral heterostructure. Here, the diffraction spots originating from $Bi_2Te_3$ (core and third layer) and $Sb_2Te_3$ (middle layer) are overlapped because of the almost identical lattice parameters of $Bi_2Te_3$ and $Sb_2Te_3$. A high-magnification BF-TEM image was captured from the edge of the heterostructure to apprehend the three-layered lateral configuration [Figure 3(c)]. The aberration-corrected ($C_s$-corrected) high-angle annular dark-field scanning TEM (HAADF-STEM) technique was employed to visualize further and distinctly recognize the trilateral heterojunctions. Figure 3(d) shows the in-plane STEM-HAADF image of the corresponding trilateral heterostructures on hBN; a single trilateral heterostructure is depicted in Figure 3(e). The HAADF image contains specific Z-dependent information in terms of contrast. In this mode, the darker contrast indicates low Z (atomic number) or thin material, whereas the brighter contrast represents high Z or thick material. Hence, Figures 3(d-e) unveil the formation of trilateral heterojunctions. The core triangular structure corresponds to $Bi_2Te_3$, surrounded by $Sb_2Te_3$ as the second layer, and the outer layer is $Bi_2Te_3$. The in-plane atomic-resolution HAADF-STEM micrograph of the $Bi_2Te_3$–$Sb_2Te_3$ and $Sb_2Te_3$–$Bi_2Te_3$ heterointerfaces taken along [0001] is depicted in Figure 3(e). As evident, no unwanted or amorphous layer was visible between the $Sb_2Te_3$–$Bi_2Te_3$ and $Bi_2Te_3$–$Sb_2Te_3$ junctions, revealing that they were atomically connected. In Figure 3(f), the bottom, middle, and top insets are the fast Fourier transform (FFT) patterns taken from the $Bi_2Te_3$ (core), $Sb_2Te_3$ (middle layer), and $Bi_2Te_3$ (last layer) regions in Figure 3(e), respectively. These patterns affirm the hexagonal symmetry of the components, along with the pronounced epitaxial coalition: $\{11\text{-}20\}_{BT(core)} \parallel \{11\text{-}20\}_{ST(2nd\ layer)} \parallel \{11\text{-}20\}_{BT(3rd\ layer)}$.

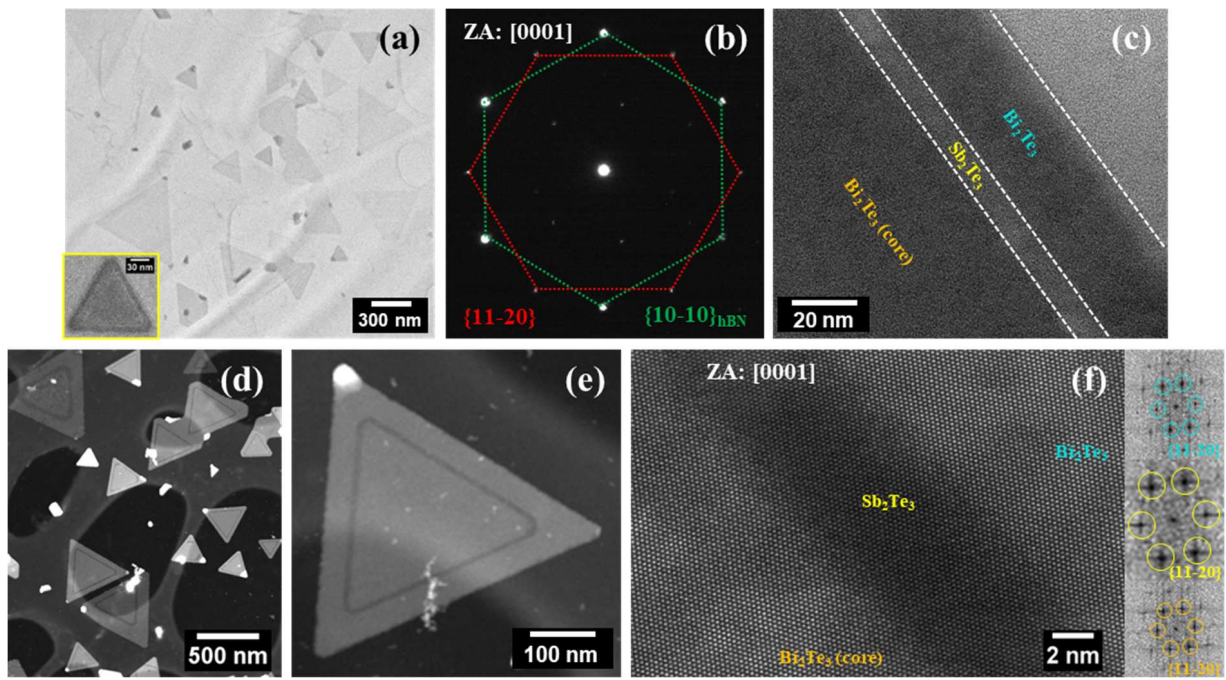

**Figure 3.** (a) Low-magnification plan-view BF-TEM micrographs of the heterostructures grown on single-crystalline hBN. (b) SAED pattern taken from a single heterostructure (shown in the inset of (a)) along the [0001] zone axis. (c) High-magnification BF-TEM image of a trilateral heterostructure (edge side). Low-magnification plan-view HAADF-STEM image of the as-grown (d) trilateral heterostructures and (e) a single trilateral heterostructure; (f) Atomic-resolution HAADF-STEM (aberration-corrected) micrograph of the $Bi_2Te_3$–$Sb_2Te_3$–$Bi_2Te_3$ interfaces (ZA: [0001]); top, middle, and bottom insets are the FFT patterns taken from the $Bi_2Te_3$, $Sb_2Te_3$, and $Bi_2Te_3$ (core) regions shown in Figure (e), respectively.

To survey the character of as-synthesized $Bi_2Te_3$–$Sb_2Te_3$–$Bi_2Te_3$ lateral heterostructures, we have executed elemental analyses in the STEM mode. Figure 4(a) represents the STEM-HAADF micrograph of a single trilateral heterostructure to examine compositional variation. A line-scan EDS was carried out across the junctions (see Figure 4(a)) to verify the lateral growth. From Figure 4(b), it is evident that the Bi signals appear only from the core area and third layer, whereas Sb appears exclusively in the second layer. Notably, no Sb signal was detected in the core and third layer, and Bi was absent from the intermediate region, confirming sharp compositional boundaries. In addition, point EDS measurements at three distinct locations (i.e., the (1) core, (2) second layer, and (3) third layer) on another trilateral heterostructure were also executed, and the corresponding results are featured in the inset of Figure 4(c). Evidently, Bi and Te elements were detected at positions (1) and (3), while Sb and Te signals were observed at point (2), with no cross-contamination. These observations indeed strongly confirm the successful synthesis and chemical integrity of the trilateral heterostructures.

Further, we have studied the plasmonic properties of the $Bi_2Te_3$–$Sb_2Te_3$–$Bi_2Te$ lateral heterostructures via EELS at the low-loss region. Figures 4(d) and 4(e) show two different single trilateral heterostructures with different thicknesses. Comparatively, the heterostructure shown in Figure 4(d) is thinner than that shown in Figure 4(e), as the latter annular dark field (ADF) image indicates a higher contrast than that of the former. The low-loss EELS spectra, obtained from the highlighted regions (blue rectangular boxes) on the respective ADF micrographs, are presented in the middle insets of Figures 4(d-e). We subtracted the background intensity to characterize the plasmonic behavior, as shown in these spectra. These background-subtracted spectra revealed the surface plasmon of both heterostructures in the ultraviolet–visible (UV–Vis) region (~ 3–5 eV). The low-loss EELS spectra from the core and edge do not show any substantial difference in both cases. Plausibly, the thicknesses of the heterostructures were sufficiently thin for the out-of-the-plane confinement effect to observe the surface plasmons even at the core. However, the peaks at approximately 6 and 8 eV in both the low-loss EELS spectra can be attributed to the plasmons of hBN (growth template) and lacey C (of the TEM grid), respectively.[38] The peak at ~ 18 eV in both cases is due to the bulk plasmons of the heterostructures.[39-41]

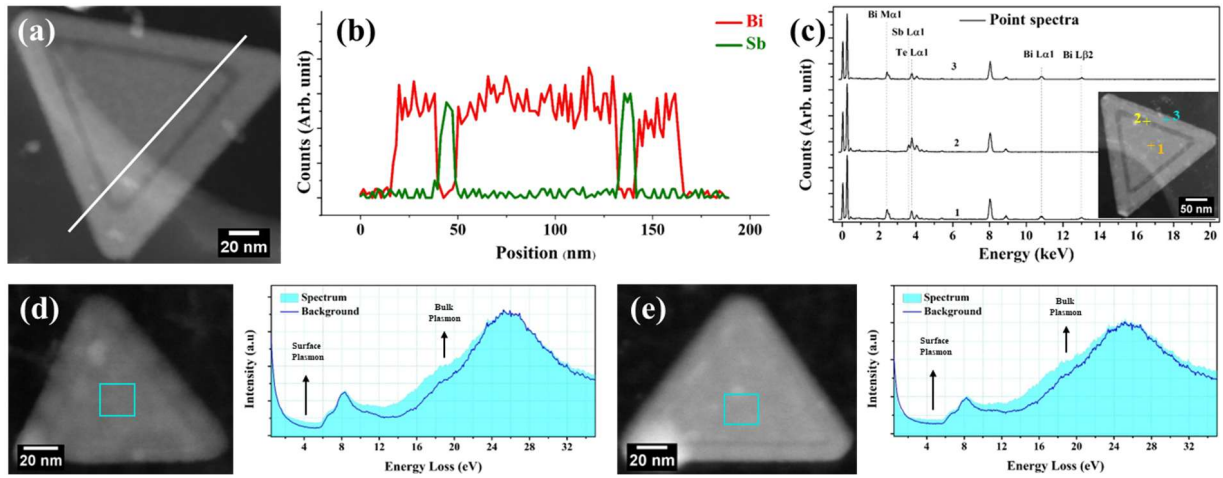

**Figure 4.** (a) STEM-HAADF micrograph of a single heterostructure; (b) line profiles of Bi and Sb elements; (c) elemental point spectra at three different locations on a single heterostructure, highlighted in the STEM-HAADF micrograph (inset). (d-e) STEM-ADF images of the trilateral heterostructures and their respective EELS spectra (corresponding areas on the heterostructures are highlighted).

In our growth strategy, the widths of individual layers, particularly the 2$^{nd}$ and 3$^{rd}$ layers, can be tuned by adjusting the growth durations at their particular stages. The relevant experimental findings are presented in the supporting information (SI). Notably, our technique can enable the reduction of layer widths to ~2–3 nm, making it suitable for investigating quantum confinement effects. For example, Figure S2 in the SI shows a $Bi_2Te_3$–$Sb_2Te_3$–$Bi_2Te_3$ trilateral heterostructure with an ~3 nm width of the $Sb_2Te_3$ layer atomically sandwiched between $Bi_2Te_3$ layers. Using the same synthetic protocol, high-quality periodic lateral (multilateral) heterostructures can also be fabricated. As demonstrated in Figure S3 (SI), the $Sb_2Te_3$–$Bi_2Te_3$–$Sb_2Te_3$–$Bi_2Te_3$ lateral heterostructures have been successfully realized. Our findings can endeavor new opportunities towards developing novel complex multilateral heterojunction and lateral superlattice fabrications. These architectures may enable exploration of quantum wells and quantum barriers in group V–VI TIs, with potential applications in transistors, spintronics, quantum computing, and related fields.

## 2.1 Theoretical Calculation

Here, we have considered the few quintuple layers (QLs) (2- to 4-QLs) $Bi_2Te_3$-$Sb_2Te_3$ lateral periodic heterostructures and explored electronic states and the topological properties by first-principles calculations. A schematic illustration of the superlattice model is shown in Figure 5(a). As illustrated in Figure 5(b), stacked QLs of $Bi_2Te_3$-$Sb_2Te_3$ lateral superlattices exhibit a slight decrement in band gaps as the domain width increases. However, the band gaps are not significantly decreased within ~30 Å of periodic domain widths for all the QLs considered here. Because of the quantum confinement,[42,43] band gaps in superlattices are much higher than ~ 0.1 eV of homogeneous few-QLs $Bi_2Te_3$[44] and $Sb_2Te_3$.[45] Since we considered periodic structures having very narrow widths of the components, the quantum size confinement effect might be strong along the lateral direction, which governs no significant change in bandgap along the domain width. Besides, a noticeable reduction in bandgap with increasing QL number is also observed in Figure 5(b. The QLs are weakly coupled vertically via van der Waals forces, while strong covalent bonding exists laterally between the constituents of lateral periodic structures.[21] These bonding characteristics likely influence the electronic structure trends noted in our calculations.

In addition, we also evaluated the topological nature of the superlattice structure. Therefore, $Z_2$ topological invariants $\nu_0$ are calculated within the framework of the Fu-Kane theory.[46] The transitions from topological insulator ($\nu_0$=1) to trivial insulator ($\nu_0$=0) at the specific domain width for each number of QLs, as shown in Figure 5(b), are observed. We found that only one band inversion is observed at the Γ-point of the Brillouin zone for all 2- to 4-QLs $Bi_2Te_3$-$Sb_2Te_3$ lateral periodic heterostructures in $\nu_0$=1. These transitions likely stem from the interaction between the surface states on vicinal heterojunctions, like vertical interlayer interactions in few-QLs $Bi_2Se_3$ or $Bi_2Te_3$.[46] However, the underlying mechanism behind the transition from a nontrivial state to a trivial one is unclear, and further theoretical studies are required.

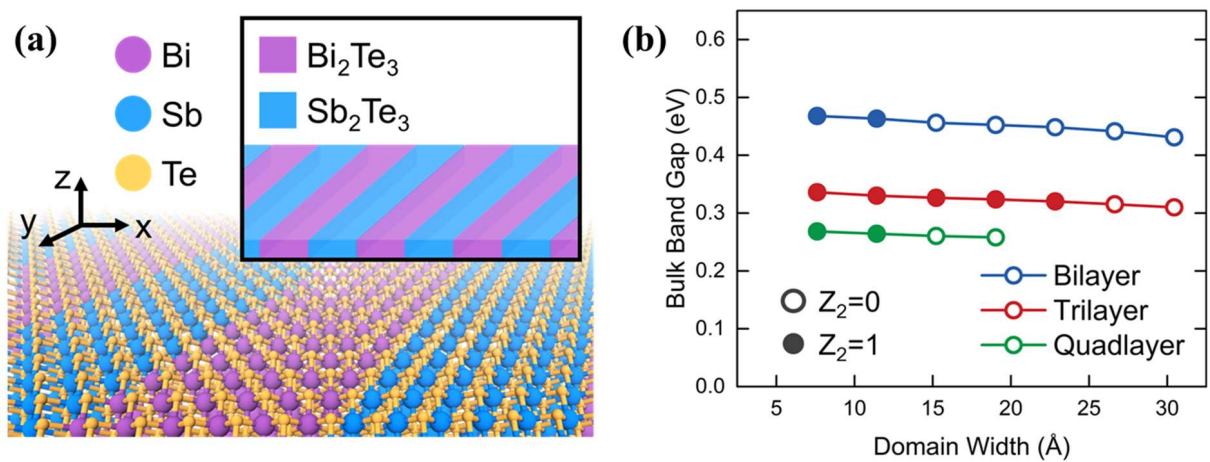

**Figure 5.** (a) Schematic illustration and perspective view of the $Bi_2Te_3$–$Sb_2Te_3$ lateral periodic heterostructure with equivalent domain widths. Bi, Sb, and Te atoms are shown as purple, blue, and orange spheres, respectively. (b) Calculated energy band gaps at the Γ-point in the bulk region of the $Bi_2Te_3$–$Sb_2Te_3$ lateral periodic heterostructures; empty and filled circles indicate the $Z_2$ topological invariants $v_0$ of 0 and 1, respectively.

## 3. Conclusion

In summary, we have auspiciously devised a novel synthesis strategy to engineer $Bi_2Te_3$–$Sb_2Te_3$–$Bi_2Te_3$ and $Sb_2Te_3$–$Bi_2Te_3$–$Sb_2Te_3$–$Bi_2Te_3$ lateral heterostructures on single-crystalline hBN templates inside a UHV chamber. The lateral heterostructures were observed to grow epitaxially on hBN; the {11-20} of the heterostructures was oriented at 30° (< ±4°) with respect to {10-10}$_{hBN}$, indicating {10-10}$_{heterostructure}$ ∥ {10-10}$_{hBN}$. The TEM analyses also confirmed high-epitaxial alignment among the components; {11-20} of $Bi_2Te_3$ (core), $Sb_2Te_3$, and $Bi_2Te_3$ (outer) were parallel to each other. The EELS studies revealed a distinct plasmonic response (surface plasmons) within the UV–Vis range in the $Bi_2Te_3$–$Sb_2Te_3$–$Bi_2Te_3$ lateral heterostructures. These findings suggest their strong potential for UV–Vis plasmonic applications. In addition, we have investigated the electronic states of few-layer multilateral heterostructures by DFT calculations. The band gaps of lateral superlattices were found to be higher than those of the homogeneous few-QLs $Bi_2Te_3$ and $Sb_2Te_3$, attributed to the quantum confinement effect in the few-QLs superlattice. Our theoretical analyses of topological properties for all 2- to 4-QLs $Bi_2Te_3$-$Sb_2Te_3$ lateral periodic heterostructures exhibited that only one band inversion is spotted at the Γ-point of the Brillouin zone in $v_0$=1. These transitions could be due to the interaction between the surface states at vicinal heterojunctions. We presume our findings can be implemented to commence new and promising research in 2D TIs with multilateral heterojunctions.

## 4. Experimental Section/Methods

This study realized a controlled, high-quality, lateral heteroepitaxial growth for $Bi_2Te_3$-$Sb_2Te_3$-$Bi_2Te_3$ lateral heterostructures using a multi-step growth process executed inside a custom-built UHV-MBE system.[4,5,47,48] A schematic of the MBE setup used to fabricate lateral heterostructures is portrayed in Figure 1(a). Commercially available single-crystalline hBN crystals were mechanically exfoliated onto ~ 300 nm thick $SiO_2$/Si substrates to serve as the growth templates. These growth templates were then transferred into the main UHV growth chamber and thermally cleaned at 450 °C for 30 min under UHV before growth. High-purity elemental sources (Bi, 99.999 %; Sb, 99.9999 %; and Te, 99.9999 %) were provided from their respective effusion cells at different stages of the growth process (Figure 1(b)). First, faceted $Bi_2Te_3$ structures (core) were fabricated on the hBN via a two-step growth process by delivering enough nucleation centers for $Sb_2Te_3$ in the second stage.[5] In the next step, with the supply of Sb and Te, $Sb_2Te_3$ started nucleating at the edge of the faceted $Bi_2Te_3$, forming a lateral epitaxial layer.[5] In the last stage, Bi and Te fluxes were again provided to grow laterally at the edge sites of $Sb_2Te_3$. Each step was performed at optimized temperatures to suppress Bi–Sb alloy formation.[5] A Te-rich environment was maintained throughout, with the Te flux 16–18 times higher than the Bi and Sb fluxes (maintained at ~0.02 Å/sec), as measured using a quartz crystal microbalance kept near the substrate.

Subsequently, electron microscopy and atomic force microscopy (AFM) techniques were exploited to study the surface morphology and structural properties of the as-prepared materials. The surface morphologies of the as-grown structures were investigated by field emission gun-based scanning electron microscope (FEG-SEM) with 5 kV electrons (MERLIN Compact, ZEISS). The height profile/thickness of the as-prepared lateral heterostructures was monitored using an AFM system (Park NX-10, Park Atomic Force Microscope) in the Non-Contact mode. The structural characterization and elemental analyses of the heterostructures were carried out using field-emission based transmission electron microscopy (TEM) with an acceleration voltage of 200 kV (analytical TEM JEM-2100F, JEOL Ltd, and Tecnai F20, FEI), equipped with energy-dispersive X-ray spectroscopy (EDS) detectors. The atomic-resolution imaging of the heterostructures was conducted with an aberration-corrected ($C_s$-corrected) monochromated scanning transmission electron microscope (STEM) operated at 200 kV (Themis Z, Thermo Fisher). The electron energy loss spectroscopy (EELS) study was performed using the Themis Z system (Thermo Fisher) with 80 kV. For the plane-view TEM observations, lateral heterostructures/hBN samples on TEM grids (ultrathin carbon film on a lacey carbon-supported Cu grid) were prepared using a wet transfer method.[4,5] First, poly(methyl methacrylate) (PMMA) was spin-coated on the heterostructures/hBN/~300 nm $SiO_2$/Si specimens; then they were treated with buffered oxide etch (BOE) to etch out $SiO_2$. Afterward, the heterostructures/hBN samples coated with the PMMA support film were transferred onto TEM grids, assisted by the wet transfer process. Finally, the heterostructures/hBN samples on the TEM grids were obtained by dissolving the PMMA layer in acetone, followed by rinsing in isopropyl alcohol and drying naturally. The TEM specimens were plasma cleaned before TEM measurements.

## 4.1. Computational Details

The electronic structures and topological characteristics were investigated by first-principles calculation within the density functional theory (DFT) framework as implemented in the VASP code.[49] For constructing the few-QLs $Bi_2Te_3$–$Sb_2Te_3$ lateral periodic heterostructure, we employed the experimental lattice parameter of $Bi_2Te_3$.[5] The distance between QLs follows that of bulk $Bi_2Te_3$. For $Sb_2Te_3$, the in-plane lattice mismatch of ~3% is negligible.[5] The vacuum slab of ~15 Å was added in a perpendicular direction to eliminate spurious interlayer interaction in the periodic cell scheme. The exchange-correlation energy functional parameterized by Perdew-Burke-Ernzerhof (PBE) based on the generalized gradient approximation was used.[50] The kinetic cutoff energy was set to 600 eV for the plane-wave basis set. The 4×16 in-plane Γ-centered k-point mesh was used for the Brillouin zone integration. Electronic optimizations were performed until the total energy differences reached below $10^{-8}$ eV. Spin-orbit coupling was considered to explore topological properties. Irreducible representations of electronic states obtained by DFT calculation were analyzed using the Irvsp code.[51]

## Supporting Information

Supporting Information is available from the author.

## Acknowledgment

This work was financially supported by the Global Research Laboratory Program (2015K1A1A2033332) and the National Research Foundation of Korea (NRF2021R1A5A1032996).

# Supporting Information

**Figure S1:**

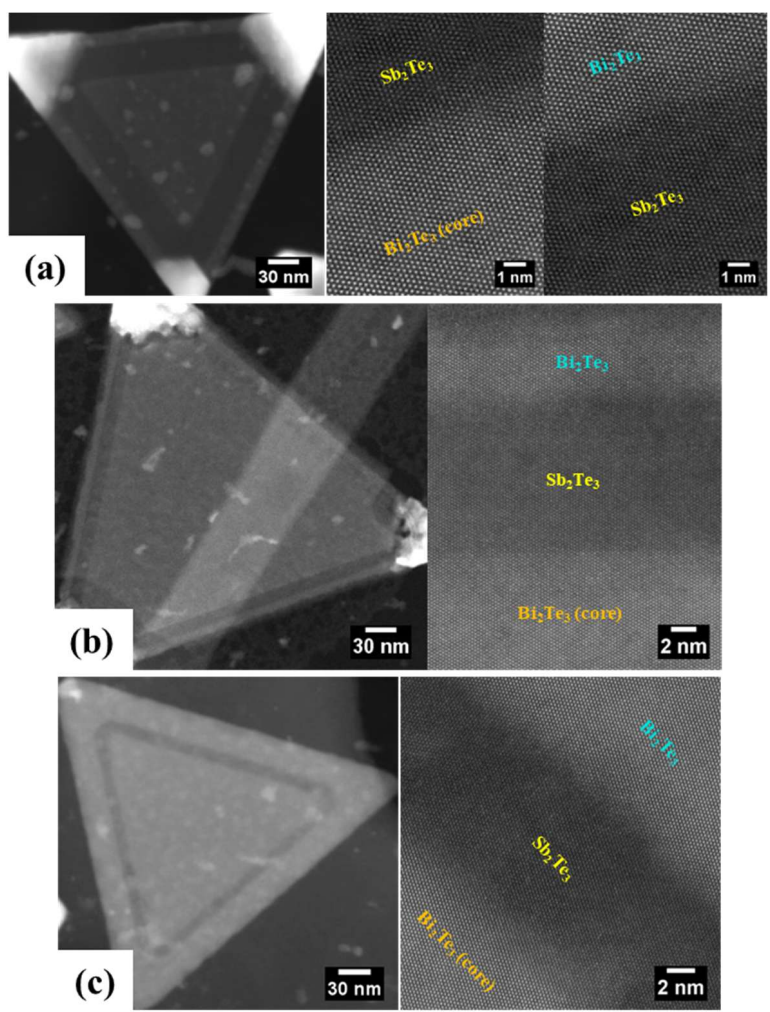

**Figure S1.** (a-c) Low-magnification and atomic-resolution STEM-HAADF micrographs of the trilateral heterostructures with different widths of the second and third layers.

**Description:**

In this work, we have varied the widths of the layers (concentrating on the second and third layers) by changing the growth durations at the respective growth stages of these layers. Figures S1 (a-c) exhibit the low-magnification STEM-HAADF micrographs of the trilateral heterostructures with different widths of the second and third layers. However, the extra brighter parts at the edges shown in Figures S1 (a-c) are due to the excess of Te. The corresponding atomic-resolution STEM-HAADF images of the heterointerfaces of the respective categories are shown in Figure S1. These results validate the successful formation of the $Bi_2Te_3$–$Sb_2Te_3$ and $Sb_2Te_3$–$Bi_2Te_3$ lateral epitaxial heterojunctions with no unwanted or amorphous layer between these heterojunctions.

**Figure S2:**

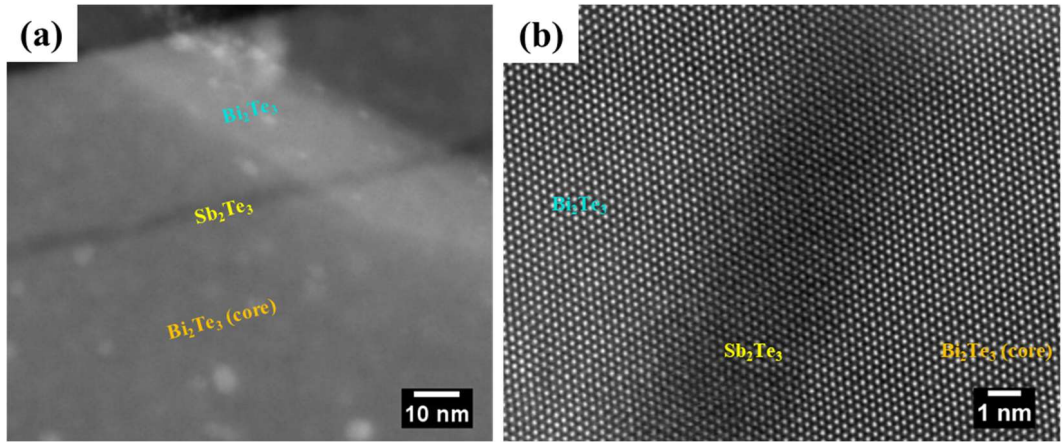

**Figure S2.** Low-magnification and atomic-resolution STEM-HAADF images of the trilateral heterojunctions, having the width of second layer ($Sb_2Te_3$) ~ 3 nm.

**Description:**

The widths of the layers were varied by changing the growth durations at the particular growth stages of these layers and were even reduced to 2–3 nm. Figure S2(a) and S2(b) show the low-magnification and atomic-resolution $C_s$-corrected STEM-HAADF images of the highly crystalline $Bi_2Te_3$–$Sb_2Te_3$–$Bi_2Te_3$ lateral heterointerface, possessing a ~ 3-nm-thick $Sb_2Te_3$ layer sandwiched between the $Bi_2Te_3$ layers, respectively. From the atomic-resolution images, the components appear epitaxially and atomically connected without any unwanted or amorphous interfacial layer. The trilateral heterostructures with $Sb_2Te_3$ (widths < 5 nm)[32] sandwiched between the $Bi_2Te_3$ layers can exhibit quantum size phenomena. Here, the quantum effect was observed in the bandgap of superlattice structures by theoretical calculations. To the best of our knowledge, the impact of the quantum confinement in TI materials has not been much investigated experimentally.[42,43] Our system could lead to a new promising research direction toward the realization of quantum confinement.

**Figure S3:**

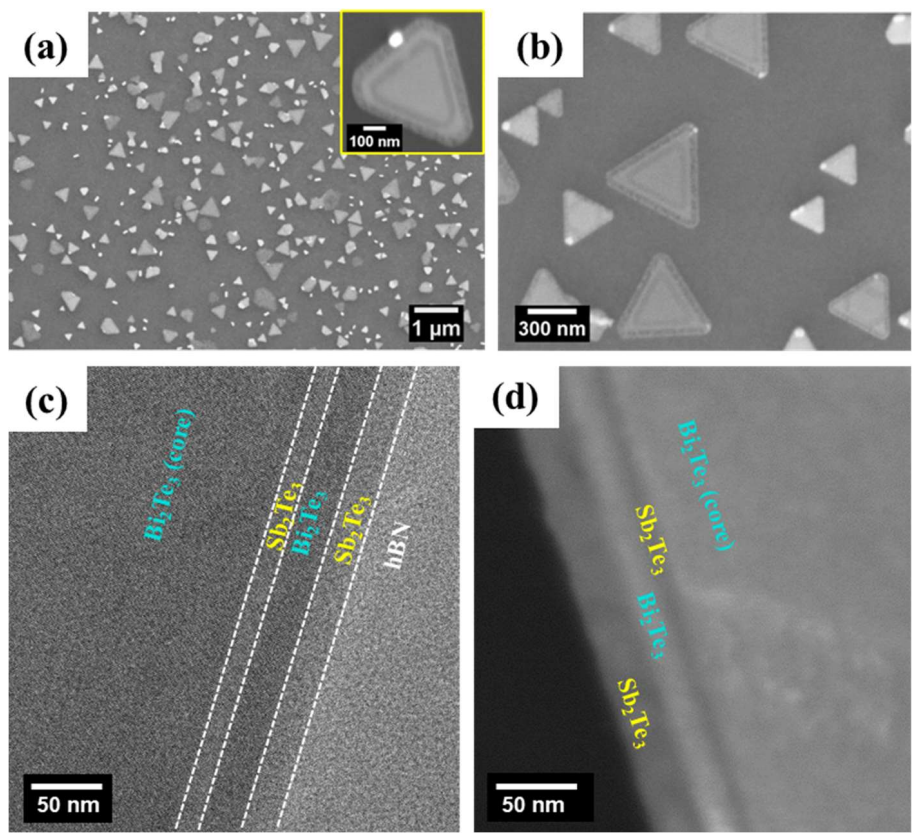

**Figure S3.** (a-b) SEM images of the as-grown faceted $Sb_2Te_3$–$Bi_2Te_3$–$Sb_2Te_3$–$Bi_2Te_3$ lateral heterostructures; inset of (a) shows a single heterostructure. Low-magnification (c) BF-TEM and (d) STEM images of the multilateral heterostructures to establish 4-layered growth in the lateral direction.

**Description:**

Figures S3(a) and S3(b) represent the SEM micrographs of the as-grown faceted $Sb_2Te_3$–$Bi_2Te_3$–$Sb_2Te_3$–$Bi_2Te_3$ lateral heterostructures. A single heterostructure is framed in the inset of Figure S3(a). These results validate the four-layer growth of the multilateral heterostructures with large-scale spatial uniformity. Figures S3(c) and S3(d) represent the low-magnification BF-TEM and STEM-HAADF images, respectively, showing the edges of the multilateral heterostructures. These images further establish the four-layered growth of the epitaxial lateral heterostructures, proposing our synthesis approach in the formation of complex multilateral heterojunctions based on group V-VI TIs.

**Table of Content**

Controlled, high-quality, lateral heteroepitaxial growth of $Bi_2Te_3$–$Sb_2Te_3$–$Bi_2Te_3$ and periodic structure ($Bi_2Te_3$–$Sb_2Te_3$–$Bi_2Te_3$–$Sb_2Te_3$) through novel *in-situ* multiple growth steps at different stages inside an ultra-high vacuum chamber, with atomically stitched lateral heterojunctions and without any amorphous or unwanted phases at the junctions

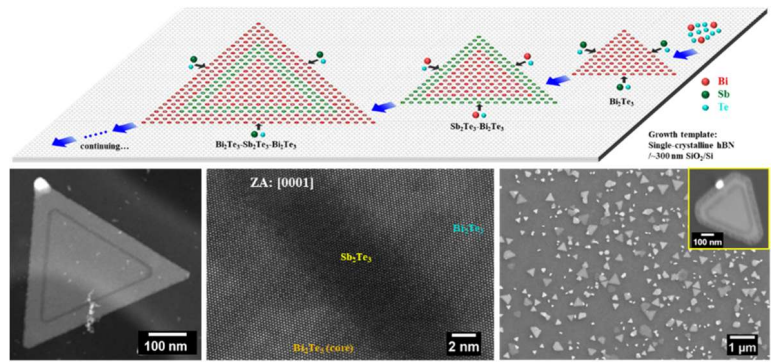